\documentstyle[preprint,prb,aps,epsf]{revtex}
\tightenlines 
\begin{document}


\title{SUPERCONDUCTIVITY IN A MESOSCOPIC DOUBLE SQUARE LOOP:
EFFECT OF IMPERFECTIONS}

\author{V. M. Fomin\cite{A1},
J. T. Devreese
\cite{A3}
}

\address{Theoretische Fysica van de Vaste Stof,
Universiteit Antwerpen (U.I.A.),
Universiteitsplein 1, B-2610 Antwerpen, Belgi\"e}

\author{V. Bruyndoncx, V. V. Moshchalkov}

\address{Laboratorium voor Vaste-Stoffysica en Magnetisme,
Katholieke Universiteit Leuven,
Celestijnenlaan 200 D, B-3001 Leuven, Belgi\"e}

\date{\today}
\maketitle

\begin{abstract}
We have generalized the network approach to include the effects of
short-range imperfections in order to analyze recent experiments on
mesoscopic superconducting double loops.
The presence of  weakly scattering imperfections causes gaps in the phase boundary
$B(T)$ or $\Phi(T)$ for certain intervals of $T$, which depend on the
magnetic flux penetrating each loop.
This is accompanied by a critical temperature $T_c(\Phi)$,
showing a smooth transition between symmetric and antisymmetric states.
When the scattering strength of imperfections increases beyond a certain
limit, gaps in the phase boundary $T_c(B)$ or $T_c(\Phi)$ appear
for values of magnetic flux lying in intervals around
half-integer $\Phi_0=hc/2e$.
The critical temperature corresponding to these values of magnetic flux
is determined mainly by imperfections in the central branch.
The calculated phase boundary is in good agreement with experiment.
\end{abstract}



Early experiments \cite{L57,C59} have revealed
that the effect of nonmagnetic impurities on the transition
temperatures of bulk superconductors is very small. The critical temperature,
$T_c$, changes by about 1\% for 1\% concentration of impurities.
An interpretation of these observations was first proposed by
Anderson\cite{A59}. The only effect of the impurities is to change
the energy of a free electron to eigenenergies
determined by those impurities. This modifies the density of states
in the integral equation for $T_c$.
Therefore, scattering by non-magnetic impurities only slightly
changes $T_c$ in bulk superconductors\cite{R65}.
Investigations extending Anderson's work have been performed
for multiband superconductors using the
Abrikosov-Gor'kov approach\cite{GM97}. An interesting
opportunity to intensify the
effect of the imperfections occurs in mesoscopic superconducting
structures where the { confined condensate is}
much more sensitive to the { action}
of impurities than in bulk structures.

Recently, the onset of the superconducting state has been studied
\cite{vital,M98}
in different mesoscopic structures of Al, comprising lines, dots, loops,
double loops, microladders etc., with sizes smaller than the coherence
length $\xi(T)$.
Refs. \onlinecite{F97,F98} show that experimentally observed phase boundaries
for square loops with two attached leads
are in excellent agreement with calculations based on
Ginzburg-Landau (GL) theory \cite{GL50,LL75}.

In this present letter, we analyze the case of a  superconducting
In this present letter, we analyze the case of a  superconducting
mesoscopic double square loop {(see the inset to Fig. 1a)}
where experiments\cite{vital} have revealed
the phase boundary shown in Fig. 1a. Using the
micronet approach\cite{DG,fink,A83} for a superconducting double loop,
we obtain a phase boundary (cf. Ref. \onlinecite{M98})
which consists of the intersecting parabolas
shown in Fig. 1b. One set (with minima at integer values of the
magnetic flux $\Phi$ through one loop in units of the
magnetic flux quantum $\Phi_0\equiv hc/2e$: $\Phi/\Phi_0$)
depends on the magnetic flux quanta penetrating each loop with
$L=0, 1, 2, \ldots$. The other set (with minima at half-integer values of
$\Phi/\Phi_0$) depends on odd numbers of magnetic flux quanta
penetrating the double loop as a whole with $L=1/2, 3/2, \ldots$.
Since the critical temperature
{corresponds to} the lowest
Landau level
$E_{LLL}(\Phi)$, the way the parabolas intersect means that the
minimum energy encounters a shift from one branch to another at certain
values of magnetic flux. Moreover, since the derivative of the
lowest Landau level with respect to magnetic flux is proportional to the
persistent current (cf. Ref. \onlinecite{IM97} Eq. (4.5)) we arrive at the
following paradox:
At the intersections, the left and right derivatives of
$E_{LLL}(\Phi)$ are different, the persistent current
has a discontinuity and its value is consequently indefinite.
In order to resolve this paradox, an analogy between
superconduct{ing} loops and semiconduct{ing} quantum rings
can be exploited.
In such rings, in the presence of impurities and
an Aharonov-Bohm magnetic field, a crossing is known to change into an
anti-crossing. In this case, the gaps between the different
eigenenergies as a function of the
magnetic flux widen and hence the degeneracy of states, which contribute
to the lowest level, is raised (see Ref. \onlinecite{WFK94}). In this
letter we demonstrate that this also applies to the superconducting
mesoscopic double square loop.

\ifpreprintsty\iffirstfig
\global\firstfigfalse
\fi\fi
\begin{figure}[h]
\protect\centerline{\epsfbox{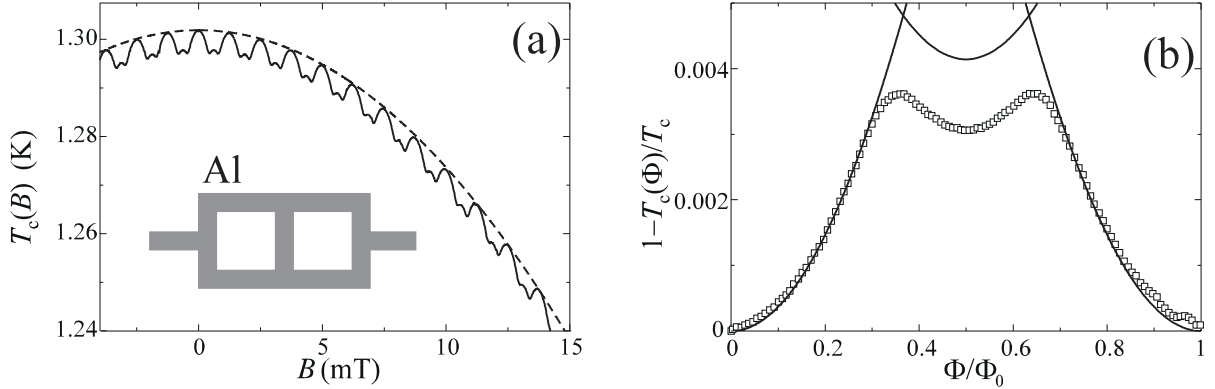}}
\caption{Experimental phase boundaries for a double loop of Al
(shown in the inset)
with (panel a) and without (panel b) a parabolic
background (the dashed line in panel a) due to the finite width of the
loop.
When transforming the
phase boundary to the plane $T_c(\Phi)/T_c$ versus $\Phi/\Phi_0$, the value
$Q$=$1302$ nm is taken in order to ensure that the maxima
of $T_c(\Phi)/T_c$ are at integer values of $\Phi/\Phi_0$.
Theoretical phase boundaries (solid lines in panel b)
calculated for a perfect double loop with $Q$=$1302$ nm
and $\xi_0=$ 128 nm.
}
\end{figure}

The observed phase boundary for a superconducting mesoscopic double
square {Al} loop is given in Fig. 1a. After subtraction of the
parabolic background, {related} to the finite width, $w= 130$ nm,
of the stripes (dashed line in Fig. 1a),
{at least} 12
{practically identical periods are seen}. One such period is plotted in Fig. 1b
(dots).
From the period of the oscillations of the phase boundary with respect to
the magnetic field, $\Delta B = 1.24$ mT, an effective loop side
length $Q = 1.3$ $\mu$m is obtained. This is close to the average
loop size. For further experimental details, the reader is
referred to Ref. \onlinecite{vital}.
In order to understand the observed ``anti-crossing'' of the different
elements of the experimentally observed phase boundary and, in particular,
the smooth shape of the $T_c(\Phi)$ minima,
we consider a superconducting mesoscopic double square loop
with imperfections. These imperfections may be introduced during
the fabrication of the mesoscopic structures.
Most probably one of the sources of such imperfections is the
inhomogeneity of the geometrical superconducting lines written by
{e}-beam lithography.

Within the framework of the GL approach, the presence of
imperfections {in} a superconducting structure
may be modelled by a spatial inhomogeneity in the parameters $a$ and $b$
in the GL equation:
\begin{eqnarray}
&&\frac{1}{2m}\left(-i\hbar\nabla-\frac{2e}{c}\mbox{\bf A}({\bf r})\right)^{2}
\psi({\bf r})+
a({\bf r})\psi({\bf r})
+b({\bf r}) \mid\psi({\bf r})\mid^{2}\psi({\bf r})=0.
\label{2}
\end{eqnarray}
Near the phase boundary, where the order parameter
$\psi({\bf r})$ is small, the system is adequately described by
the linearized GL equation:
\begin{eqnarray}
\frac{1}{2m}\left(-i\hbar\nabla-\frac{2e}{c}\mbox{\bf A}({\bf r})\right)^{2}
\psi({\bf r})+ a({\bf r})\psi({\bf r})=0.
\label{3}
\end{eqnarray}
Moreover, the magnetic field may be assumed to be equal to the applied
magnetic field. The vector potential $\bf A$ of the uniform field
${\bf B}\parallel {\bf e}_z$ is taken in the symmetric gauge.

The presence of imperfections{,} localized in the loop around several points
${\bf r}_s$ over a distance which is much smaller than the coherence length
or the typical loop size (``short range imperfections''){,}
can be modelled by the following function:
\begin{eqnarray}
a({\bf r}) = a + \sum_sV_s\delta({\bf r}-{\bf r}_s),
\label{4}\end{eqnarray}
where $a$ is the GL {coefficient} of the substance{,}
and the magnitudes
$V_s$ are determined by specific characteristics of the imperfections.
Eq. (\ref{3}) then
becomes similar
to the Schr\"odinger equation for a particle of mass
$m$ and charge $2e$ in the potential field described by the scalar form
$\sum_sV_s\delta({\bf r}-{\bf r}_s)$ and by the vector potential
$\mbox{\bf A}({\bf r})$, the quantity $-a$ playing the role of the
energy.

Short-range imperfections are assumed to be present in all three branches
of the loop at the points characterized by the coordinates
$Q_s$ with $s=L,M,R$ as shown in Fig. 2.
Furthermore, taking $\xi(T)\equiv\xi_0/\sqrt{(1-T/T_c)}$ as a unit of length,
we obtain the linearized GL equation in terms of dimensionless
coordinates:
\begin{eqnarray}
&&\Bigg\{(i \nabla_x+{2\pi \over \Phi_0}A_x)^2 +
(i \nabla_y+{2\pi \over \Phi_0} A_y)^2
+ \sum_{s=L,M,R}{\tilde V}_s\delta(y-Q_s)-1 \Bigg\}\Psi = 0.
\label{3b}
\end{eqnarray}
It should be noted that the dimensionless scattering magnitude
${\tilde V}_s = 2mV_s\xi(T)/\hbar^2 = C_s/\sqrt{(1-T/T_c)}$
is tempe\-ra\-ture dependent.

\begin{figure}[h]
\protect\centerline{\epsfbox{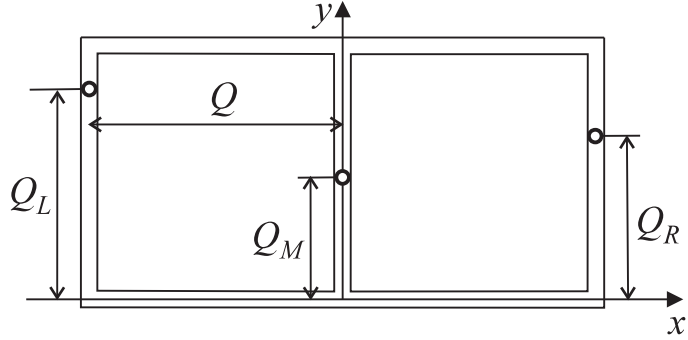}}
\caption{A model configuration of imperfections (open circles)
in the double loop.}
\end{figure}

We have solved Eq. (\ref{3b}) following the micronet
approach.  A new conceptual feature compared with
Refs.~\onlinecite{DG,fink,A83} is the use of
{\it additional nodal points} {at}
the positions of {the} imperfections.
The presence of these imperfections implies an additional
condition, which can be derived in the following way.
In the vicinity of the point with $y=Q_s$, Eq. (\ref{3b})
takes the form
\begin{equation}
(i \nabla_y+{2\pi \over \Phi_0} A_y)^2\Psi -\Psi =
- {\tilde V}_s\delta(y-Q_s)\Psi.
\label{3a}
\end{equation}
We integrate both sides of this equation over $y$ from
$Q_s-\epsilon$ to $Q_s+\epsilon$, $\epsilon \to +0$. Taking into account
continuity of the order parameter,
$\Psi(Q_s+\epsilon)=\Psi(Q_s-\epsilon)$, we obtain the additional
constraint \begin{equation}
\left.{d\Psi\over dy}\right\vert_{y=Q_s+\epsilon}
-\left.{d\Psi\over dy}\right\vert_{y=Q_s-\epsilon}=
{\tilde V}_s\Psi(y=Q_s)
\label{dif}
\end{equation}
for the derivatives
to the left and to the right from the new nodal point $y=Q_s$.
Applying Eq. (\ref{dif}) to a single loop of
length ${\cal L}$ with one imperfection leads to the following secular
equation
\begin{eqnarray}
\cos(\varphi) = \cos({\cal L}) + {1\over 2} {\tilde V}_s\sin({\cal L}),
\label{one}\end{eqnarray}
{where $\varphi = {2\pi \Phi /\Phi_0}$}.

Analyzing the onset of superconductivity in a superconducting ring
with a lateral arm of length $\ell$, de Gennes \cite{DG} derived an equation
which differs from Eq. (\ref{one}) by the substitution  ${\tilde V}_s =
-\tan(\ell)/2$. Thus, from the point of view of our present analysis,
a lateral arm may be considered of as a kind of imperfection
in a ring.

After some lengthy, but straightforward algebra, we obtain the secular
equation which determines the phase boundary:
\widetext
\begin{eqnarray}
\cos^2\varphi + \cos\varphi {D_L+D_R\over 2D_M}
-{D_LD_R\over 4}\left[\prod_{i=1}^2\left({N_{iM}\over D_M}+{N_{iL}\over D_L} +
{N_{iR}\over D_R}\right) - {1\over D_M^2}-{1\over D_L^2}-{1\over D_R^2}\right]
-{1\over 2}=0
\label{sec}
\end{eqnarray}
with
\begin{eqnarray}
&&N_{1s} = \cos 3Q+ {\tilde V}_s \sin(2Q-Q_s)\cos(Q+Q_s),
N_{2s} = \cos 3Q+ {\tilde V}_s \cos(2Q-Q_s)\sin(Q+Q_s),\nonumber\\
&&D_{s} = \sin 3Q+ {\tilde V}_s \sin(2Q-Q_s)\sin(Q+Q_s);\nonumber\\
&&N_{1M} = \cos Q+ {\tilde V}_M \sin(Q-Q_M)\cos(Q_M),
N_{2M} = \cos Q+ {\tilde V}_M \cos(Q-Q_M)\sin(Q+Q_M),\nonumber\\
&&D_{M} = \sin Q+ {\tilde V}_M \sin(2Q-Q_M)\sin(Q+Q_M),
\end{eqnarray}
where $s=L,R$.

\narrowtext
In Eq. (\ref{sec}), $\varphi = {2\pi \Phi /\Phi_0}$
with $\Phi$, the magnetic flux through each of the loops.
Here{,} we recall that lengths in the above equations are expressed in units
of $\xi(T)$ and are therefore functions of {the} temperature.
Consequently, the
secular equation establishes a relation between the magnetic
flux $\Phi$ and the temperature $T$.


The phase boundaries obtained by solving Eq. (\ref{sec}),
with an imperfection  in only one branch of the {double} loop, are
shown in Figs. 3 to {5} ($L$ in Figs. 3 and 4, $M$ in Fig. 5).

\begin{figure}[h]
\protect\centerline{\epsfbox{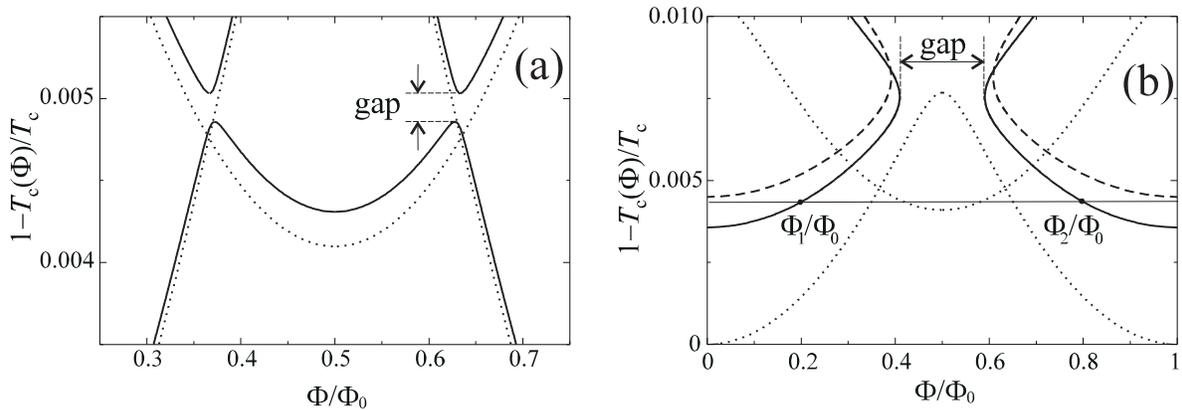}}
\caption{Theoretical phase boundaries for a double loop
with an imperfection in one side branch at $Q_L=0.5Q$ for
$Q$=$1302$ nm
and
$\xi_0=$ 128 nm. {(a)} The dimensionless magnitude
$C_L$= $0$ and $0.01$
corresponds to dotted and solid
lines, respectively.
{(b)} The dimensionless magnitude
$C_L$= $0, 0.5, 1.0$
corresponds to dotted, solid and dashed lines.
The values $\Phi_1$ and $\Phi_2$ are discussed in the text.
}
\end{figure}

As is seen in Fig. 3{a}, at small values $C_s$,
an ``energy'' gap forms between solutions corresponding to different
numbers of
magnetic flux quanta penetrating each loop ($L=0,1/2,1,...$).
The resultant phase boundary indicates that a continuous change takes place
from a {\it symmetric} superconducting order parameter at integer values of
$\Phi/\Phi_0$ to an {\it antisymmetric} state at half-integer values of the
relative magnetic flux.
It is also worthy of note that the presence of
imperfections {slightly} diminishes the critical temperature of
{the} double loop at zero magnetic field.

When increasing $C_s$ above a certain
limit, the pattern of phase boundaries changes dramatically. Gaps appear
in certain {flux} intervals (``flux gaps'')
around half-integer $\Phi/\Phi_0$ values.
This behavior is illustrated {by the curves} in Fig. {3b}.
For a given $T<T_c$,
a superconducting state exists
when $\Phi$ ranges from zero to
a certain value of $\Phi_1$, after which the sample turns
{in}to the normal state.
With a further increase of magnetic flux
{along a horizontal straight line shown in Fig. 3b}, the sample remains in the normal
state until a value of $\Phi_2$ is reached, at which the sample
becomes {again} superconducting.
This demonstrates a re-entrant {behavior}
as a function of field. {[When approaching in Fig. 3b the points where
the phase boundary $\Phi(T)$ has its extrema and, consequently, the derivatives
$\partial T/\partial \Phi$ would diverge, the superconducting state
apparently becomes unstable.]}
It should be noted
that the existence of such a regime,
where the system is normal {in} a certain {flux} interval,
was reported by de Gennes \cite{DG} for
a single superconducting ring with a lateral arm.

The trend of lowering of
the critical temperature of a double loop at zero magnetic field
with increasing $C_s$ is
clearly seen by comparison of the curves in Fig. {3b} which refer
to different values of $V_L$.

\begin{figure}[h]
\protect\centerline{\epsfbox{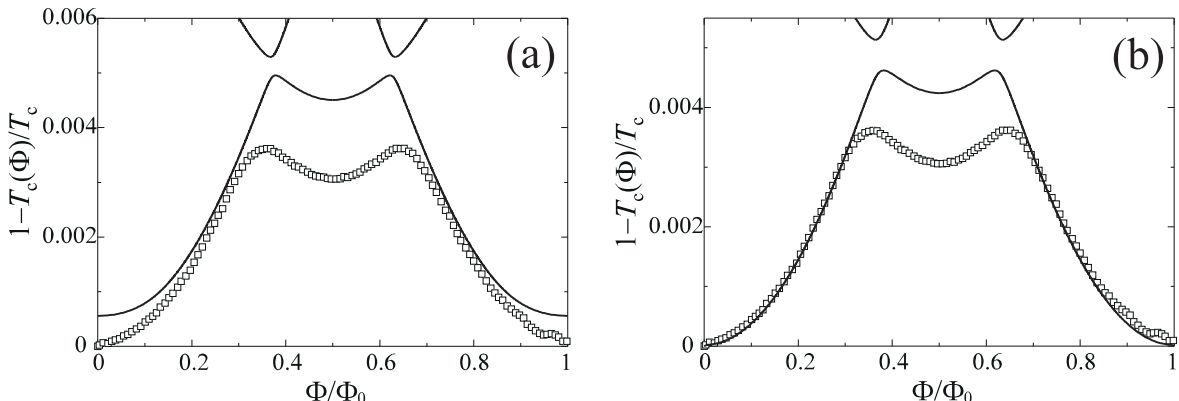}}
\caption{Theoretical phase boundaries (solid lines) for a double loop
with an imperfection in one side branch
at $Q_L=0.5Q$ for $Q$=$1302$ nm and
$\xi_0=$ 128 nm. The dimensionless magnitude
is $C_L$= $0.02.$
Results obtained without and with
renormalization of temperature are given in panels a and b, respectively.
Experimental data are shown with points.
}
\end{figure}

In regions adjacent to integer values of
$\Phi/\Phi_0$, {a} good
agreement with experiment
of the calculated phase boundary $T_c(\Phi)$,
is achieved for the
zero-temperature coherence length
$\xi_0$ = 128 nm. This value is in accordance with previous estimates
(see  Ref. \onlinecite{MGS95}).  The resulting set of phase
boundaries is shown in Fig. {4}a.  For comparison with {the}
experimental data,
it is necessary to renormalize
the temperature scale, taking as a unit temperature the specific critical
value of a loop with imperfections {(see}
Fig. {4}b).

In Fig. {5}, a plot of the phase boundary is shown for a double loop
with an imperfection in the middle branch ($M$ in Fig. 2).
It is clear that by increasing $V_M$, one shifts a minimum
of the $1-T_c(\Phi)/T_c$ curve (using a renormalized temperature),
at half-integer values of $\Phi/\Phi_0$, to higher values, without modifying
those parts of the phase boundary close to the integer values of
$\Phi/\Phi_0$.

\begin{figure}[h]
\protect\centerline{\epsfbox{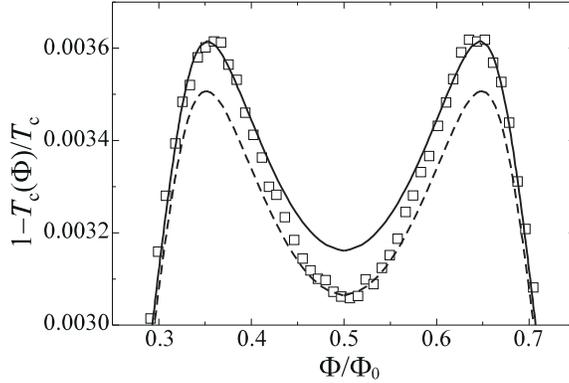}}
\caption{Theoretical phase boundaries for a double loop
with an imperfection in the middle branch at $Q_M=0.1Q$ for
$Q$=$1302$ nm and
$\xi_0=$ 128 nm. The dimensionless magnitude
$C_M$= $0.11, 0.125$
corresponds to dashed and solid lines.
Results are obtained with renormalization of temperature.
Experimental data are shown with points.
}
\end{figure}

The best agreement between the calculated phase boundary and
experimental data is achieved
for a configuration where imperfections are present
on {\it all three} branches of the double loop ($L,M,R$ in Fig. 2).
The corresponding {curves} are shown in  Fig. {6}.
The main observation which follows from these figures
is that imperfections in the {\it central branch} play a decisive role in
determining the critical temperature of a double loop at the half-integer
values of
$\Phi/\Phi_0$.

\begin{figure}[h]
\protect\centerline{\epsfbox{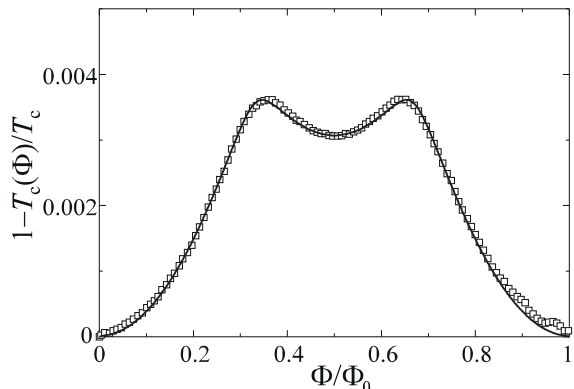}}
\caption{Theoretical phase boundaries (solid line) for a double loop
with imperfections in all three branches at $Q_L=Q_R=Q_M=0.5Q$ for
$Q$=$1302$ nm and
$\xi_0=$ 128 nm. The dimensionless magnitudes are:
$C_L$= $-C_R$ = $0.017$
and $C_M$= $0.090$.
Results are obtained with renormalization of temperature.
Experimental data are shown with points.
}
\end{figure}


In conclusion, we have generalized the network approach to include the
effects
of short-range imperfections. The presence of weakly scattering
imperfections leads to the formation of
gaps between solutions corresponding to integer and half-integer numbers
of magnetic flux quanta penetrating each loop.
The phase boundary is characterized by
a smooth transition between symmetric and antisymmetric states.
For imperfections with relatively large magnitudes,
{\it gaps in the phase boundary
$T_c(B)$ or $T_c(\Phi)$ appear} when
the magnetic flux lies in intervals around half-integer values.
The critical temperature at the half-integer values of the relative magnetic
flux has been shown to be determined mainly by imperfections in the central
branch. The calculated phase boundary for a mesoscopic double square loop is
in good agreement with experiment.

{\it Acknowledgments.} -
We thank V. N. Gladilin for fruitful interactions.
This work has been supported by
the Interuniversitaire Attractiepolen --- Belgische Staat, Diensten van de
Eerste Minister --- Wetenschappelijke,
technische en culturele Aangelegenheden;
the F.W.O.-V. projects Nos. G.0287.95, G.0232.96, G.0306.00
W.O.G. WO.025.99N (Belgium), and the ESF Programme VORTEX.


\begin{references}
\bibitem[*]{A1}
Also at:
Technische Universiteit Eindhoven, P. O. Box 513,
5600 MB Eindhoven, The Netherlands.
Permanent address: Department of Theoretical Physics,
State University of Moldova,
str. A. Mateevici, 60,
MD-2009 Kishinev, Republic of Moldova.

\bibitem[\sharp]{A3}
Also at: Universiteit Antwerpen (RUCA), Groenenborgerlaan 171,
B-2020 Antwerpen, Belgi\"e and
Technische Universiteit Eindhoven, P. O. Box 513,
5600 MB Eindhoven, The Netherlands.

\bibitem{L57} E.A. Linton, B. Serin, and M. Zucker, J. Phys. Chem. Solids
{\bf 3}, 165 (1957).

\bibitem{C59} G. Chanin, E.A. Linton, and B. Serin, Phys. Rev. {\bf 114},
719 (1959).

\bibitem{A59} P. Anderson, J. Phys. Chem. Solids
{\bf 11}, 26 (1959).

\bibitem{R65} G. Rickayzen, {\em Theory of Superconductivity},
John Wiley \&Sons, New York, 1965.

\bibitem{GM97} A.A. Golubov and I.I. Mazin, Phys. Rev. B {\bf 55},
15146 (1997).

\bibitem{vital}
V. Bruyndoncx, C. Strunk, V. V. Moshchalkov, C. Van Haesendonck
and Y. Bruynseraede, Europhys. Lett. 36, 449 (1996).

\bibitem{M98} V. V. Moshchalkov, V. Bruyndoncx, L. Van Look,
M. J. Van Bael, Y. Bruynseraede, and A. Tonomura,
in: H.~S.~Nalwa, Ed., {\it Handbook of Nanostructured Materials
and Nanotechnology}, Vol. 3, Chap. 9, pp. 451-525,
Academic Press, San Diego, 1999.

\bibitem{F97} V. M. Fomin, V. R. Misko, J. T. Devreese, and V.~V.~Moshchalkov,
Solid State Communications {\bf 101}, 303 (1997).

\bibitem{F98} V. M. Fomin, V. R. Misko, J. T. Devreese, and V. V. Moshchalkov,
Phys. Rev, B {\bf 58}, 11703 (1998).

\bibitem{GL50}
 V.~L.~Ginzburg and L.~D.~Landau, Zh. Eksp. i Teor. Fiz. {\bf 20},
 1064 (1950).

\bibitem{LL75}
 L.~D.~Landau and E.~M.~Lifshitz, {\it Course of Theoretical Physics,}
 Vol. 9 ({\it Statistical Physics,} Vol. 2), Pergamon, Oxford, 1989.

\bibitem{DG} P.-G. de Gennes, C. R. Acad. Sci. Paris {\bf 292}, II - 279
(1981).

\bibitem{fink}
H.~J.~Fink, A.~L\'{o}pez, and R.~Maynard,
Phys. Rev. B {\bf 26}, 5237 (1982).

\bibitem{A83}
 S. Alexander, Phys. Rev. B\ {\bf 27}, 1541 (1983).

\bibitem{IM97}
 Y.~Imry, {\it Introduction to Mesoscopic Physics,}
 Oxford University Press, Oxford, 1997.

\bibitem{WFK94} L. Wendler, V. M. Fomin, and A. A. Krokhin,
Phys. Rev. B\ {\bf 50}, 4642 (1994).

\bibitem{MGS95}
 V.~V.~Moshchalkov, L.~Gielen, C.~Strunk, R.~Jonckheere, X.~Qiu,
 C.~Van~Haesendonck, and Y.~Bruynseraede, Nature {\bf 373}, 319
 (1995).

\end{references}
\end{document}